\documentclass[
nobibnotes,aps,pre,showkeys,
12pt]{revtex4-2}
\usepackage{hyperref}
\usepackage{graphics}
\usepackage{subcaption}
\usepackage{graphicx}
\usepackage{color}
\usepackage{amsmath}
\usepackage{amssymb}
\usepackage{bm}
\usepackage{float}
\usepackage{epsfig}
\usepackage{epstopdf}
\usepackage{color,graphics}
\usepackage{mathtools}

\begin{document}

\title{Electron Holes in a Regularized Kappa Background}

\author{Fernando Haas}
\affiliation{Instituto de F\'{\i}sica, Universidade Federal do Rio Grande do Sul, Av. Bento Gon\c{c}alves 9500, 91501-970 Porto Alegre, RS, Brasil}

\author{Horst Fichtner and Klaus Scherer}
\affiliation{Institut f\"ur Theoretische Physik, Lehrstuhl IV: Plasma-Astroteilchenphysik, Ruhr-Universit\"at Bochum, D-44780 Bochum, Germany;  Research Department, Plasmas with Complex Interactions, Ruhr-Universit\"at Bochum, D-44780 Bochum, Germany}

\begin{abstract}
The pseudopotential method is used to derive electron hole structures in a suprathermal plasma having a regularized $\kappa$ probability  distribution function background. The regularized character allows the exploration of small $\kappa$ values beyond the standard suprathermal case, for which $\kappa > 3/2$ is a necessary condition. We have found the nonlinear dispersion relation yielding the amplitude of the electrostatic potential in terms of the remaining parameters, in particular the drift velocity, the wavenumber and the spectral index. Periodic, solitary wave, drifting and non-drifting solutions have been identified. In the linear limit, the dispersion relation yields generalized Langmuir and electron acoustic plasma modes. Standard electron hole structures are regained in the $\kappa \gg 1$ limit. 
\end{abstract}

\maketitle

\section{Introduction}
The phenomenon of so-called electron holes in a plasma has received growing attention 
in the recent past specially due to recent spacecraft observation of 
such structures, see, e.g., \citep{Steinvall2019a, Steinvall2019b} 
General reviews on electron holes can be found e. g. in \citep{Luque2005, Eliasson2006}. For the application to space plasmas a quantitative treatment of electron
holes should take into account the presence of a suprathermal, i.e.\ non-Maxwellian 
background plasma. This was already pointed out in \citep{Schamel2015, Schamel2023} and carried 
out in \citep{Haas2021, w1, w2, w3}. In \citep{Haas2021} the Maxwellian description of the 
trapped (hole) and untrapped (background) electron populations was substituted by one 
with a so-called standard kappa distribution (SKD).

The SKD is a simple generalization of a Maxwellian that was originally introduced by 
\citep{Olbert1968}  to describe non-Maxwellian power-law distributions of suprathermal 
plasma species, which are frequently observed in the solar wind \citep{Lazar2017}  
and are formed via the interaction of the solar wind particles with the plasma turbulence
\citep[e.g.,][]{Ma-Summers-1998, Yoon2014, Yoon-2018} preventing a relaxation to a
Maxwellian or bi-Maxwellian. Since then the SKD
has been applied successfully to numerous space plasma and 
laboratory scenarios. Along with these successes also various limitations of the SKD 
were identified: it exhibits diverging velocity moments, a positive lower limit of
allowed kappa parameter values ($\kappa > 3/2$), and a non-extensive entropy (for a
recent overview see \citep{Lazar2021}). In addition, two types of SKDs were
identified, namely the original one introduced by Olbert \citep{Olbert1968} with a 
prescribed reference speed and a modified one that can be traced to Matsumoto
\citep{Matsumoto1972} with a temperature equal to that of the associated Maxwellian, 
and it was demonstrated \citep{Lazar2016} that care has to be taken in
selecting one of those for a given physical system. The kappa distribution was
proposed in \citep{k1}; 
extensive discussion on the
different kappa distributions can be found in \citep{k2, k4}. 
Besides these principal limitations of SKDs, there is also an observational one: 
SKDs do not allow to describe velocity distributions which are harder than $v^{-5}$.
However, distributions with harder tails are actually observed, see, for example,
\citep{Gloeckler-etal-2012}. At the same time, these measurements also reveal that
kappa values near two or below are frequently observed. This can also be seen for solar
wind electrons, see, e.g.\ \citep{Pierrard-etal-2022}. Such low values of kappa imply
unphysical features of the SKD as is discussed in \citep{Scherer2019}.
Another example are solar energetic particles, see, e.g., \citep{Oka-etal-2013}.  
Kappa values as low as 1.63 and two are also obtained for particle distributions
in the outer heliosphere \citep[e.g.,][]{Heerikhuisen-etal-2008, Zirnstein-etal-2017}.
Finally, SKDs are not consistent with exponential cut-offs of observed power-law distributions
of suprathermal proton in the solar wind \citep{Fisk-Gloeckler-2012}.
 
All of these complications in employing the SKD can be avoided when one uses the 
\textit{regularized kappa distribution} (RKD) introduced non-relativistically in
\citep{Scherer2017} and for the relativistic case in \citep{HanThanh2022}.  The RKD 
exhibits an exponential cut-off of the power at high velocities. Such cut-off is a 
result of the fact that any acceleration process can only occur on a finite spatial scale
and a finite time scale. Consequently, such power law cannot extend to infinity (as in
the case of the standard kappa distribution) but must cut-off.
The main purpose of the present work is to adopt a regularized version of
the SKD and to analyze the consequences. The RKD
particularly removes all divergences in the theory and moves the lower limit 
for the kappa parameter to zero \citep{Scherer2019}. Both features have consequences for correspondingly 
described physical systems: in \citep{Yoon2014} it was demonstrated that an `infrared 
catastrophe' is avoided when using the RKD instead of the SKD and in \citep{Liu2020} it
was shown that extending the range of kappa values to zero broadens the possible 
properties of solitary ion acoustic waves in a plasma with RKD electrons.
 Here the reference value of $\kappa$ is adopted according to Eq. (\ref{tit}) for the SKD.

Since also the first generalization of the analytical treatment of electron holes in
an equilibrium plasma to a suprathermal plasma was achieved by employing the SKD 
\citep{Haas2021}, the same constraints remain: not all moments of the velocity 
distribution functions exist and kappa has to be greater than 3/2, thereby potentially 
preventing the study of a physically interesting regime  because harder velocity 
distributions are observed, see, e.g.\ \citep{Gloeckler-etal-2012, Pierrard-etal-2022} 
and were associated with observations of various solitary waves \citep{Vasko2017}. 
Therefore, the present work 
revisits the quantitative treatment of electron holes in a suprathermal plasma, where
the electron velocity distribution is described with the RKD. 

The structure of the paper is as follows: in section~II the one-dimensional RKD is
defined, in section~III various dimensionless variables are introduced, in section~IV 
the method of the  pseudopotential is applied and in section~V special solutions of
the resulting Poisson equation are derived. After an analysis of the corresponding  
dispersion relation in section~VI for homogeneous trapped electrons distributions, the final section~VII contains the conclusions of 
the study.

\section{One-dimensional regularized $\kappa$ distribution}
The starting point \citep{Scherer2019, Liu2020} is the three-dimensional isotropic regularized $kappa$ distribution (RKD), 
\begin{equation}
f_3({\bf u}) = \frac{n_0}{(\pi\kappa\theta^2)^{3/2}\,U\left(\frac{3}{2},\frac{3}{2}-\kappa,\alpha^2\kappa\right)}\,\left(1 + \frac{u^2}{\kappa\theta^2}\right)^{-\kappa-1}\,\exp\left(-\,\frac{\alpha^2 u^2}{\theta^2}\right) \,,
\end{equation}
where $n_0$ is the equilibrium electrons number density, $\kappa > 0$ is the spectral index, $\theta$ is a reference speed, $U$ is a Kummer function of the second kind (or Tricomi function) described in \citep{Scherer2019, Liu2020,  Abramowitz1972}, ${\bf u}$ is the velocity vector with $u = |{\bf u}|$ and $\alpha \geq 0$ is the cutoff parameter. 

In the non-regularized limit $\alpha \rightarrow 0$ one regains the SKD 
\begin{equation}
\label{tit}
f_{3}({\bf u}) = \frac{n_0\,\Gamma(\kappa+1)}{(\pi\kappa\theta^2)^{3/2}\,\Gamma\left(\kappa - \frac{1}{2}\right)}\,\left(1 + \frac{u^2}{\kappa\theta^2}\right)^{-\kappa-1} \,, \quad \alpha \rightarrow 0 \,, 
\end{equation}
where $\Gamma$ is the gamma function, which is positive defined provided $\kappa > 1/2$. For the RKD this constraint is not imposed on $\kappa > 0$.

For the treatment of electrostatic structures it is convenient to define the one-dimensional RKD. For this purpose we use cylindrical coordinates in velocity space and write $u^2 = v^2 + w^2$, where $v$ is the component of the velocity parallel to the electric field and ${\bf w}$ contains only the perpendicular velocity components, with $w = |{\bf w}|$. In the isotropic case the one-dimensional RKD is 
\begin{eqnarray}
f(v) &=& 2\pi \int_{0}^{\infty} dw\,w\,f_{3}({\bf u}) \nonumber \\
&=& 
\nonumber 
\frac{2\pi\,n_0\,e^{-\frac{\alpha^2 v^2}{\theta^2}}}{(\pi\kappa\theta^2)^{3/2}\,U\left(\frac{3}{2},\frac{3}{2}-\kappa,\alpha^2\kappa\right)}\,\int_{0}^{\infty} dw\,w\,\,\,\left(1 + \frac{v^2 + w^2}{\kappa\theta^2}\right)^{-\kappa-1}\,\exp\left(-\,\frac{\alpha^2 w^2}{\theta^2}\right) \\ 
&=&
\label{r1}
\frac{n_0\,(\alpha^2\kappa)^{\kappa}\,e^{\alpha^2\kappa}}{(\pi\kappa\theta^2)^{1/2}\,U\left(\frac{3}{2},\frac{3}{2}-\kappa,\alpha^2\kappa\right)}\,\Gamma\left[-\kappa, \alpha^2\kappa\left(1 + \frac{v^2}{\kappa\,\theta^2}\right)\right] \,,
\end{eqnarray}
where here $\Gamma$ is the incomplete gamma function of the indicated arguments \citep{Abramowitz1972}. In other words, $f(v)$ comes from the three-dimensional version after integration over the two perpendicular velocity components. 

In the non-regularized limit $\alpha \rightarrow 0$ one regains the standard one-dimensional $\kappa$ distribution \citep{Summers, Podesta} 
\begin{equation}
f(v) = \frac{n_0\,\Gamma(\kappa)}{(\pi\kappa\theta^2)^{1/2}\,\Gamma\left(\kappa - \frac{1}{2}\right)}\,\left(1 + \frac{v^2}{\kappa\theta^2}\right)^{-\kappa} \,, \quad \alpha \rightarrow 0 \,, 
\end{equation}
which is positive definite provided $\kappa > 3/2$. 

In the treatment of electrostatic structures, to satisfy Vlasov's equation the distribution function is a function of the constants of motion. In the one-dimensional, time-independent case, the available constants of motion are given by 
\begin{equation}
\label{e4}
\epsilon = \frac{m v^2}{2} - e\phi \,, \quad \sigma = {\rm sgn}(v) \,,
\end{equation}
where $\phi = \phi(x)$ is the scalar potential, where $m$ is the electron mass and $-e$ is the electron charge. The sign of the velocity $\sigma = v/|v|$ is an additional constant of motion just in the case of  untrapped particles. The energy variable $\epsilon$ can be used to distinguish untrapped ($\epsilon > 0$) and trapped ($\epsilon < 0$) electrons. 

In analogy with \citep{Schamel1972, Schamel2015, Schamel2023} (where the background is not in the RKD form), presently one starts from Eq. (\ref{r1}) making for the untrapped part the replacement $v \rightarrow \sigma \sqrt{2\epsilon/m} + v_0$, where $v_0$ is a drift velocity, defining the distributions of untrapped and trapped electrons according to  
\begin{eqnarray} 
f = f(\epsilon,\sigma) = \frac{A\,n_0}{\theta}\!\!\!\!&\strut&\!\!\!\!\left(1 + \frac{k_0^2 \Psi}{2}\right)\,\Bigl[H(\epsilon)\,\Gamma\left(-\kappa, \alpha^2\kappa\left(1 + \frac{1}{\kappa\,\theta^2}(\sigma\sqrt{2\epsilon/m}+v_0)^2\right)\right) \nonumber \\ 
&+& H(-\epsilon)\,\Gamma\left(-\kappa, \alpha^2\kappa\left(1 + \frac{v_{0}^2}{\kappa\,\theta^2}\right)\right)\,(1 - \frac{\beta\,\epsilon}{m\,\theta^2})\Bigr] \,,
\label{f1} \\ 
A &=& \frac{(\alpha^2\kappa)^{\kappa}\,e^{\alpha^2\kappa}}{(\pi\kappa)^{1/2}\,U\left(\frac{3}{2},\frac{3}{2}-\kappa,\alpha^2\kappa\right)} \,,
\end{eqnarray}
where $H(\epsilon)$ is the Heaviside function. The quantities $k_0$ and $\Psi$ are dimensionless variables proportional respectively to the wavenumber of periodic oscillations and to the electrostatic field amplitude, as will be qualified in the following. In addition, $\beta$ is a dimensionless quantity associated to the inverse temperature  of the trapped electrons distribution. Unlike singular distributions as in \citep{Schamel2015, Schamel2023, Haas2021, kkk} here the velocity shifted hole distribution is assumed continuous at the separatrix ($\epsilon = 0$) and an analytic function of the energy for both trapped and untrapped electrons. These choices have been made in order to focus on the role of the cutoff parameter $\alpha$ instead of further aspects. 

\begin{figure}
\centering{
\includegraphics[height=8cm, width=11cm]{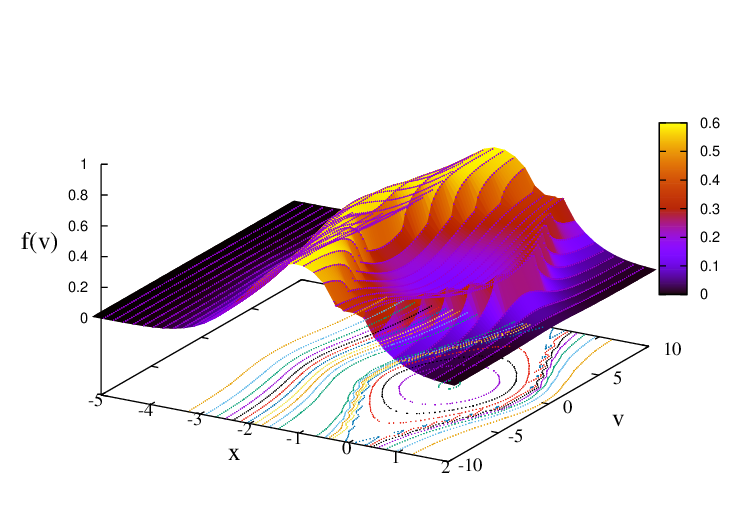}}
\vspace*{10mm}
\caption{An illustration of the Schamel distribution (in arbitrary units and for the sech-potential in Eq.(13) in \citep{Schamel1986}) for the 
values $\beta = -0.9, k_0 = 1.0, \psi = 1.0, \kappa = 0.5, \gamma  = 0.1$ of the parameters
used for the notation in \citep{Haas2021}.}
\label{fig01}
\end{figure}

In the non-regularized case, using 
\begin{equation}
(\alpha^2\kappa)^\kappa\Gamma(-\kappa,\alpha^2\kappa\,s) \rightarrow \frac{s^{-\kappa}}{\kappa} \,, \quad \alpha \rightarrow 0 \,, \quad \kappa > 0 \,,
\end{equation}
for a generic argument $s$, and 
\begin{equation}
\label{half}
U\left(\frac{3}{2}, \frac{3}{2} - \kappa, \alpha^2\kappa\right) \rightarrow \frac{\Gamma(\kappa - 1/2)}{\Gamma(\kappa + 1)} \,, \quad \alpha \rightarrow 0 \,, \quad \kappa > 1/2 \,,
\end{equation}
from Eq. (\ref{f1}) one obtains
\begin{eqnarray}
\label{e5}
f &=& \frac{n_0 \,(1 + k_0^2\Psi/2)}{(\pi\kappa\theta^2)^{1/2}}\,\frac{\Gamma(\kappa)}{\Gamma(\kappa-1/2)}\,\Bigl[H(\epsilon)\left(1 + \frac{1}{\kappa\theta^2}(\sigma \sqrt{\frac{2\epsilon}{m}} + v_0)^2\right)^{-\kappa} + \\
&+&  H(-\epsilon)\left(1 + \frac{v_0^2}{\kappa\theta^2}\right)^{-\kappa}\left(1  - \,\frac{\beta\,\epsilon}{m\,\theta^2}\right)\Bigr] \,,
\nonumber
\end{eqnarray}
which is the $\kappa$ version of Schamel's distribution that is given in its original form, e.g., in Eq. (4) in \citep{Schamel1986} and illustrated in Fig.~\ref{fig01}. A slight difference in comparison to the original formulation \citep{Schamel1986, Schamel2012} is that here the trapped electrons are described by a linear function of the energy instead of a Maxwellian function. 

Finally, the Poisson equation
\begin{equation}
\label{yy}
\frac{\partial^2\phi}{\partial x^2} = \frac{e}{\varepsilon_0}(n - n_0) \,, \quad n = n(\phi) = \int_{-\infty}^{\infty} dv\,f(\epsilon, \sigma) 
\end{equation}
is needed, where $\varepsilon_0$ is the vacuum permittivity. An uniform ionic background $n_0$ has been assumed. 

\section{Dimensionless variables}
To avoid the use of a large number of parameters, it is convenient to adopt dimensionless variables. For the RKD, it comes the question on which will be the reference speed defining the velocity rescaling. It would be tempting to consider the use of a thermal speed $v_T$ defined in terms of the averaged squared velocity, but it is a cumbersome expression containing Kummer functions, 
\begin{equation}
\label{vt}
v_{T}^2 = \frac{<u^2>}{3} = \frac{1}{3}\,\frac{\int d^{3}u\, u^2\, f_{3}({\bf u})}{\int d^{3}u\, f_{3}({\bf u})} = \frac{\kappa\theta^2}{2}\,\frac{U\left(\frac{5}{2},\frac{5}{2}-\kappa,\alpha^{2}\kappa\right)}{U\left(\frac{3}{2},\frac{3}{2}-\kappa,\alpha^{2}\kappa\right)} \,,
\end{equation}
the factor $1/3$ introduced to comply with the one-dimensional geometry.  Therefore, for the sake of simplicity, instead of the thermal speed it is indicated to consider $\theta$ as the reference speed. In this way, the rescaled variables are 
\begin{equation}
\label{re}
\tilde{x} = \frac{x}{\lambda} \,, \quad \tilde{v} = \frac{v}{\theta} \,, \quad \tilde{v_0} = \frac{v_0}{\theta} \,, \quad \tilde{\phi} = \frac{e\phi}{m\,\theta^2} \,, \quad 
\tilde{n} = \frac{n}{n_0} \,, \quad \tilde{f} = \frac{f}{n_0/\theta} \,, \quad \tilde{\epsilon} = \frac{\epsilon}{m\,\theta^2} \,, 
\end{equation}
where $\lambda = [\epsilon_0 m \theta^2/(n_0 e^2)]^{1/2}$ is a modified Debye length. 

As discussed in \citep{Lazar2016} in the non-regularized context, our standard choice of $\theta$ as a $\kappa-$independent parameter better fits a scenario with enhanced tail in velocity space. Alternatively one could choose $v_T$ from Eq. (\ref{vt}) to be $\kappa-$independent, which would be adequate for an enhanced core. 

In dimensionless variables omitting for simplicity the tildes, the one-dimensional hole RKD from Eq. (\ref{f1}) is 
\begin{eqnarray} 
f(\epsilon, \sigma) = A\!\!\!\!&\strut&\!\!\!\!\left(1 + \frac{k_0^2 \Psi}{2}\right)\,\Bigl[H(\epsilon)\,\Gamma\left(-\kappa, \alpha^2\kappa\left(1 + \frac{1}{\kappa}(\sigma\sqrt{2\epsilon}+v_0)^2\right)\right) \nonumber \\ 
&+& H(-\epsilon)\,\Gamma\left(-\kappa, \alpha^2\kappa\left(1 + \frac{v_{0}^2}{\kappa}\right)\right)\,(1 - \beta\epsilon)\Bigr] \,,
\label{fu}
\end{eqnarray}
while Poisson's equation (\ref{yy}) is  
\begin{equation}
\label{yyy}
\frac{\partial^2\phi}{\partial x^2} = n - 1 \,, \quad n = n(\phi) = \int_{-\infty}^{\infty} dv\, f(\epsilon, \sigma) \,,
\end{equation}
where $\epsilon = v^2/2 - \phi$ and $\sigma = {\rm sgn}(v)$. In the remaining, the purpose is to evaluate the number density in Eq. (\ref{yyy}) in terms of $\phi$ and to characterize the possible solutions of the Poisson's equation, specially regarding the behavior according to the parameters $\kappa, \alpha$.

\section{Pseudopotential method} 
From Eqs. (\ref{fu}) and (\ref{yyy}) one has
\begin{eqnarray}
\frac{n}{A} &=& \left(1 + \frac{k_0^2 \Psi}{2}\right)\,\Bigl[\int_{-\infty}^{-\sqrt{2\phi}}dv\,\Gamma\left(-\kappa, \alpha^2\kappa\left(1 + \frac{1}{\kappa}(\sqrt{2\epsilon}-v_0)^2\right)\right) + \nonumber \\ &+& \int_{\sqrt{2\phi}}^{\infty}dv\,\Gamma\left(-\kappa, \alpha^2\kappa\left(1 + \frac{1}{\kappa}(\sqrt{2\epsilon}+v_0)^2\right)\right) + 
\label{fuu} \\
&+& 
\Gamma\left(-\kappa, \alpha^2\kappa\left(1 + \frac{v_{0}^2}{\kappa}\right)\right)\,\int_{-\sqrt{2\phi}}^{\sqrt{2\phi}}dv\,(1 - \beta\epsilon)\Bigr] \,, 
\nonumber 
\end{eqnarray}
assuming $0 \leq \phi \leq \Psi$,  where $\Psi$ denotes the peak-to-peak amplitude of the electrostatic potential, so that at $\phi = \Psi$ one has $d\phi/dx = 0$. 

The integrals in Eq. (\ref{fuu}) for the contribution of untrapped particles can be evaluated only in the weakly nonlinear limit. Expanding the integrands in a formal power series on $\sqrt{\phi}$ 
the result is 
\begin{equation}
\label{nn}
n = 1 + \frac{k_0^2\,\Psi}{2} + a\,\phi + b\,\phi\sqrt{\phi} + {\cal O}(\phi^2) \,,
\end{equation}
 keeping the term proportional to $\Psi$ as it has the same order of magnitude of $\phi$ where 
\begin{eqnarray}
a &=& \frac{2}{\kappa\,U\left(\frac{3}{2},\frac{3}{2}-\kappa,\alpha^2\kappa\right)}\,\Bigl[U\left(\frac{1}{2},\frac{1}{2}-\kappa,\alpha^2\kappa\right) + \nonumber \\
&+& \frac{v_0}{\sqrt{\pi\,\kappa}}\,P\int_{-\infty}^{\infty}\frac{ds}{s - v_0}\,e^{-\alpha^2\,s^2}\,\left(1 + \frac{s^2}{\kappa}\right)^{-\kappa-1}\Bigr]
\label{aaa}
\end{eqnarray}
where $P$ stands for the principal value, and 
\begin{eqnarray}
b &=& \frac{4\sqrt{2}}{3}\,\beta\,A\,\Gamma\left(-\kappa, \alpha^2\kappa\left(1 + \frac{v_{0}^2}{\kappa}\right)\right) + \nonumber \\
&+& \frac{8\sqrt{2}e^{-\alpha^2 v_0^2}\left[v_0^2 + 2\alpha^2 v_0^4 + \kappa\left(-1 + 2 (1 + \alpha^2)v_0^2\right)\right]}{3\kappa^2\sqrt{\pi\kappa}\left(1 + v_0^2/\kappa\right)^{\kappa+2}U\left(\frac{3}{2},\frac{3}{2}-\kappa,\alpha^2\kappa\right)} \,.
\label{bbb}
\end{eqnarray}

It is possible to proceed in the same way to determine the average velocity $<v>$ from 
\begin{equation}
n \langle v \rangle = \int_{-\infty}^{\infty} dv\,v\,f(\epsilon,\sigma) 
\end{equation}
yielding 
\begin{equation}
\langle v \rangle = - v_0 \,(1 - a\,\phi) + {\cal O}(\phi^{3/2}) 
\end{equation}
giving a more precise meaning of $- v_0$ which is the global drift velocity only in the limit of zero field amplitude. In addition notice the trapped electrons do not contribute to the average velocity, which comes from the untrapped part only, as found from the detail of the procedure similar to Eq. (\ref{fuu}). 

Poisson's equation (\ref{yyy}) can be rewritten in terms of the  pseudopotential $V = V(\phi)$, 
\begin{equation}
\label{ppp}
\frac{d^2\phi}{dx^2} = n - 1 = - \,\frac{\partial V}{\partial\phi} \,,
\end{equation}
where 
\begin{equation}
\label{ss}
- V = \frac{k_0^2\Psi\phi}{2} + \frac{a\,\phi^2}{2} + \frac{2\,b\,\phi^2\sqrt{\phi}}{5} + {\cal O}(\phi^3) \,, 
\end{equation} 

The case where the solutions are either periodic or solitary waves requires 
\begin{enumerate}
\item \quad$V(\phi) < 0$ in the interval $0 < \phi < \Psi$;
\item \quad $V(\Psi) = 0$ \,,
\end{enumerate}
the latter implying 
\begin{equation}
\label{ndr}
k_0^2 + a + \frac{4\,b\sqrt{\Psi}}{5} = 0 \,,
\end{equation}
which allows rewriting Eq. (\ref{ss}) as 
\begin{equation}
\label{sss}
- V = \frac{k_0^2\phi}{2}(\Psi - \phi) + \frac{2\,b\,\phi^2}{5}(\sqrt{\phi} - \sqrt{\Psi}) \,,
\end{equation}
up to ${\cal O}(\phi^3)$. 

Equation (\ref{ndr}) is the nonlinear dispersion relation (NDR) of the problem, providing a relation between phase velocity $v_0$, wavenumber $k_0$ and amplitude proportional to $\Psi$, taking into account the expressions (\ref{aaa}) and (\ref{bbb}) for $a, b$. On the other hand, Eq. (\ref{ppp}) can be integrated yielding 
\begin{equation}
\label{e21}
\frac{1}{2}\left(\frac{d\phi}{dx}\right)^2 + V(\phi) = 0 \,,
\end{equation}
where the integration constant was set to zero due to property (I) and since at the potential maximum $\phi = \Psi$ the electric field is zero.   Following the usage from \citep{Schamel2015, Schamel2023, Haas2021, Schamel1972, kkk, Schamel1986, Schamel2012}, the proposed {\it Ansatz} has tailored $\Psi$ so that it is the root of $V(\phi)$ in Eq. (\ref{sss}). Otherwise, an irrelevant additive constant would be incorporated in the pseudopotential. The same applies to Eqs. (\ref{ku}) and (\ref{qdr}) below. 

\section{Special solutions}
\subsection{Periodic solutions}
 As discussed in \citep{Schamel2015, Schamel2023, Haas2021, Schamel1972, kkk, Schamel1986, Schamel2012}, the expansion of the number density in powers of $\sqrt{\phi}$ starting from an {\it Ansatz} such as in Eq. (\ref{fu}) can give periodic or localized solutions, according to specific conditions to be identified. For the sake of reference, we collect some of the known analytic solutions, remembering that of course now the coefficients are adapted to the RKD equilibrium. For localized solutions as a by-product one has decaying boundary conditions. 

The quadrature of Eq. (\ref{e21}) yields closed form solutions in special cases. 
In the linear limit,  for a small amplitude so that $\sqrt{\Psi} << k_0^2/b$, neglecting the nonlinearity term $\sim b$, one has 
\begin{equation}
\label{ku}
V = \frac{k_0^2 \phi}{2}(\phi - \Psi) \,.
\end{equation}
Then from Eq. (\ref{e21}) immediately one has 
\begin{equation}
\label{lin}
\phi = \frac{\Psi}{2}[1 + \cos(k_0\,(x - x_0))] \,.
\end{equation}
Hence it is verified that $k_0$ indeed corresponds to the wavenumber of linear oscillations with $0 \leq \phi \leq \Psi$ in this case.

\begin{figure}
\centering{
\includegraphics[width=10.5 cm]{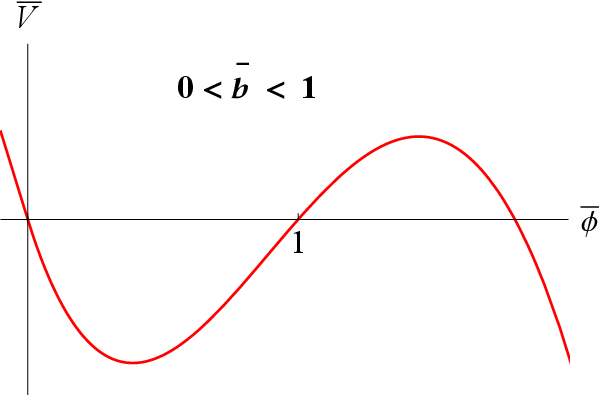}}
\caption{Rescaled pseudopotential from Eq. (\ref{qdr}) for $0 < \bar{b} < 1$.}
\label{fig02}
\end{figure}

\begin{figure}
\centering{
\includegraphics[width=10.5 cm]{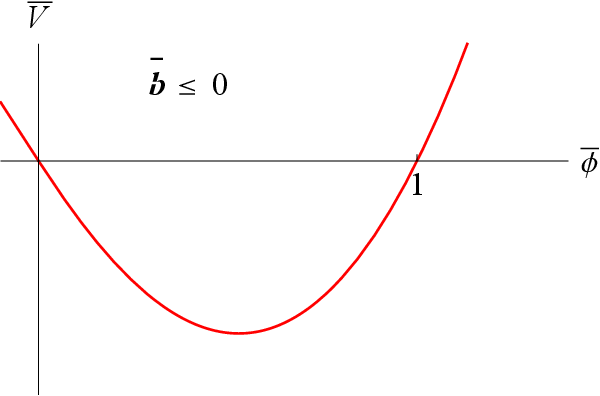}}
\caption{Rescaled  pseudopotential from Eq. (\ref{qdr}) for $\bar{b} \leq 0$.}
\label{fig03}
\end{figure}

\begin{figure}
\centering{
\includegraphics[width=10.5 cm]{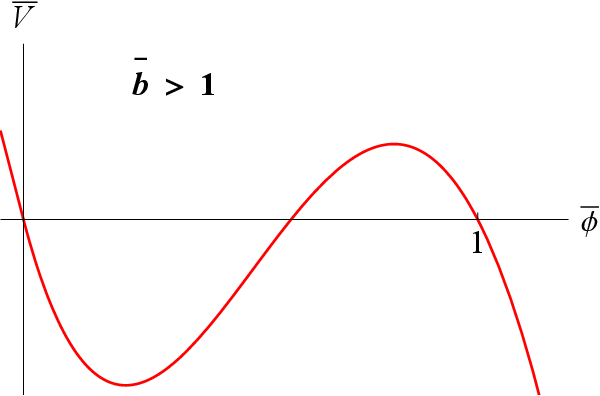}}
\caption{Rescaled  pseudopotential from Eq. (\ref{qdr}) for $\bar{b} > 1$. Periodic solutions exist in a smaller interval $0 \leq {\bar\phi} < 1$.}
\label{fig04}
\end{figure}

Assuming $k_0 \neq 0$, more insight is provided by the further rescaling
\begin{equation}
\label{fr}
{\bar\phi} = \frac{\phi}{\Psi} \,, \quad {\bar x} = k_0\,x \,, \quad {\bar V} = {\bar V}({\bar\phi}) = \frac{V}{k_0^2\Psi^2} \,, \quad {\bar b} = \frac{2\,b\,\sqrt{\Psi}}{5\,k_0^2}
\end{equation}
reduces Eq. (\ref{e21}) to 
\begin{equation}
\label{bbbb}
\frac{1}{2}\left(\frac{d{\bar\phi}}{d{\bar x}}\right)^2 + {\bar V}({\bar\phi})  = 0 \,,
\end{equation}
where 
\begin{eqnarray}
- {\bar V}({\bar\phi}) &=& \frac{\bar\phi}{2} \,(1 - {\bar\phi}) + {\bar b}\,{\bar\phi}^2\, \left(\sqrt{\bar\phi} - 1\right) \nonumber \\
&=& \frac{\bar\phi}{2}\left(1 - \sqrt{\bar\phi}\right)\,\left(1 + \sqrt{\bar\phi}- 2\,{\bar b}\,{\bar\phi}\right) \,, \label{qdr}
\end{eqnarray}
containing only one free parameter ${\bar b}$. The condition (II) for periodic or localized solutions amounts to ${\bar V}({\bar\phi}) < 0$ within the interval $0 < {\bar\phi} < 1$. In view of the factorization in Eq. (\ref{qdr}) it is easy to demonstrate the condition is always satisfied for ${\bar b} < 1$. The existence of periodic solutions such that $0 \leq {\bar\phi} \leq 1$ for ${\bar b} < 1$ comes from the shape of the rescaled  pseudopotential shown in Figs. \ref{fig02} and \ref{fig03}. The case ${\bar b} > 1$ also has periodic solutions, but with a smaller amplitude as apparent from Fig. \ref{fig04}. The physically meaningful solutions always occur for ${\bar V} < 0$ within the interval $0 < {\bar\phi} < 1$. Notice that with the further rescaling (\ref{fr}) the amplitude of oscillation is set to unity, as shown in the referred figures. The required weakly nonlinear analysis always supposes $\tilde\phi \sim \Psi \ll 1$ or, according to Eq. (\ref{re}), $e\phi/(m\theta^2) \ll 1$, where $\phi$ is the physical scalar potential.  

 The exact quadrature of Eq. (\ref{bbbb}) with all terms has been fully discussed in \citep{Schamel2012, sch}, where the pseudopotential is formally the same as in Eq. (\ref{qdr}) after rescaling. 
It is given in terms of Jacobi elliptic functions showing a periodic behavior and higher order Fourier harmonics. The present work extends these results for the case of a background RKD, with the adapted coefficients.

\begin{figure}
\centering{
\includegraphics[width=10.5 cm]{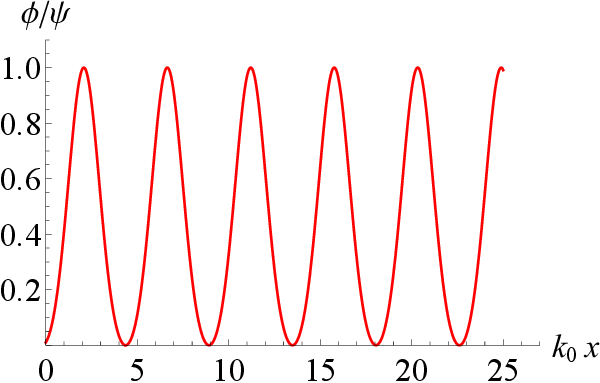}}
\caption{Numerical solution of Eq. (\ref{bbbb}) with $\bar{b} = -2, {\bar\phi}(0) = 10^{-3}$.}
\label{fig05}
\end{figure}

\begin{figure}
\centering{
\includegraphics[width=10.5 cm]{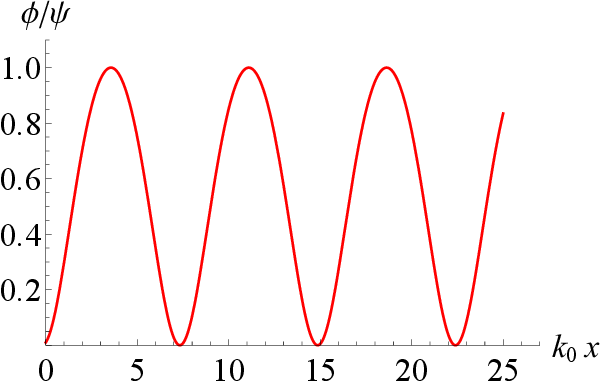}}
\caption{Numerical solution of Eq. (\ref{bbbb}) with $\bar{b} = 0.5, {\bar\phi}(0) = 10^{-3}$.}
\label{fig06}
\end{figure}

\begin{figure}
\centering{
\includegraphics[width=10.5 cm]{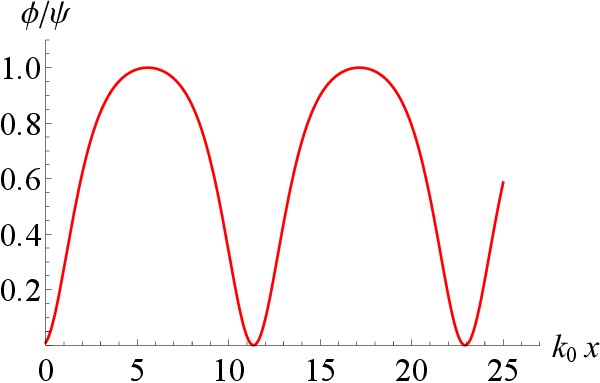}}
\caption{Numerical solution of Eq. (\ref{bbbb}) with $\bar{b} = 0.9, {\bar\phi}(0) = 10^{-3}$.}
\label{fig07}
\end{figure}

It is apparent that the control parameter ${\bar b}$ depending on several variables such as the effective trapped particles inverse temperature $\beta$ determines the qualitative aspects of the oscillatory solutions. Figures \ref{fig05} and \ref{fig06} and \ref{fig07} show in a different style how a smaller (and possibly negative) ${\bar b} < 1$ corresponds to a larger wavenumber, which is exactly $k_0$ only in the linear case.

\subsection{Localized solution with ${\bar b} = 1, k_0 \neq 0$}
\begin{figure}
\centering{
\includegraphics[width=10.5 cm]{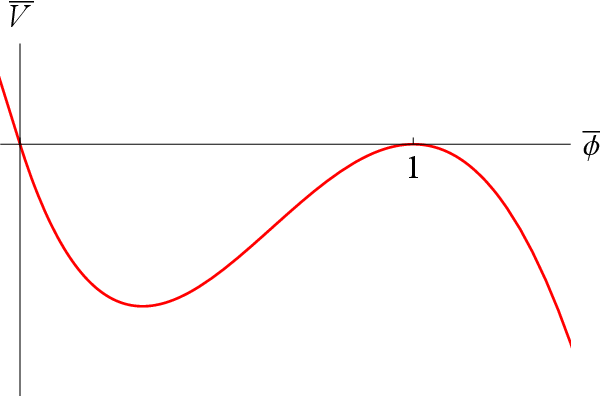}}
\caption{Rescaled  pseudopotential from Eq. (\ref{qdr}) for $\bar{b} = 1$.}
\label{fig08}
\end{figure}

\begin{figure}
\centering{
\includegraphics[width=10.5 cm]{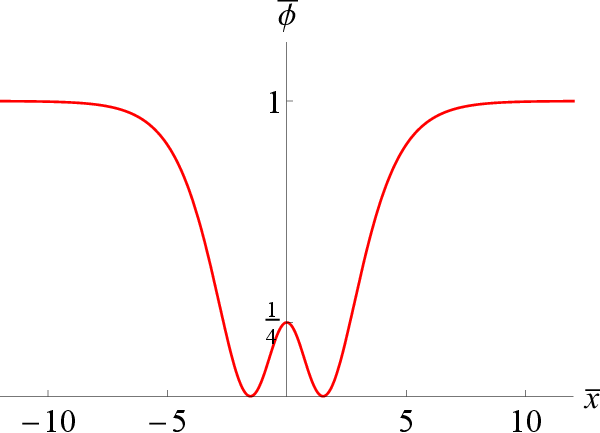}}
\caption{Rescaled electrostatic potential from Eq. (\ref{k}) for $\bar{x}_0 = 0$.}
\label{fig09}
\end{figure}

\begin{figure}
\centering{
\includegraphics[width=10.5 cm]{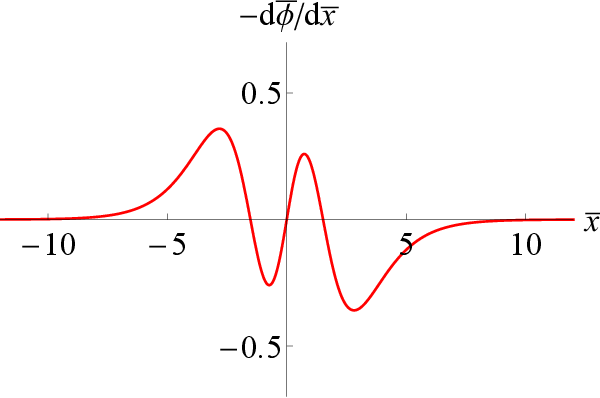}}
\caption{Rescaled electric field $-d{\bar\phi}/d{\bar x}$ where ${\bar\phi}$ is given in Eq. (\ref{k}) for $\bar{x}_0 = 0$.}
\label{fig10}
\end{figure}

The limit case ${\bar b} = 1$ with $k_0 \neq 0$ is special since then $d{\bar V}/d{\bar\phi} = 0$ at ${\bar\phi} = 1$, as shown in Fig. \ref{fig08}, yielding a localized, non-periodic solution. Moreover this case is amenable to the simple quadrature 
\begin{equation}
{\bar\phi} = \frac{1}{4}\left[1 - 3\,\tanh^{2}\left(\frac{\sqrt 3}{4}({\bar x} - {\bar x}_0)\right)\right]^2 \,,
\label{k}
\end{equation}
see Fig. \ref{fig09}. The corresponding rescaled electric field is shown in Fig. \ref{fig10}. The total electrostatic energy is finite since the integral $(1/2)\int_{-\infty}^{\infty} d{\bar x} (d{\bar\phi}/d{\bar x})^2 = 6\sqrt{3}/35$ converges.

\subsection{Solitary waves with $k_0 = 0$.}
On the other hand if $k_0 = 0$ one has 
\begin{equation}
V = \frac{2\,b\,\phi^2}{5}(\sqrt{\Psi} - \sqrt{\phi}) \,,
\end{equation}
yielding the solitary pulse
\begin{equation}
\phi = \Psi\,{\rm sech}^{4}\left[\left(\frac{-b\sqrt{\Psi}}{20}\right)^{1/2}(x - x_0)\right] \,,
\end{equation}
which is well defined everywhere provided $b < 0$, which can be attainable e.g. for sufficiently small $\beta, v_0^2$. 

\section{Dispersion relation} 
The NDR (\ref{ndr}) provides several behaviors according to the values in parameter space. For the sake of simplicity it will be 
considered the case where the trapped particle distribution is homogeneous in phase space, which amounts to the dimensionless quantity $\beta = 0$ in Eq. (\ref{e5}). This is an increasingly better approximation for small enough amplitude so that $e\Psi << m \theta^2$, yielding a relatively smaller trapped area in phase space. Clearly this limit situation does not correspond to "holes", since in this case the trapped particles are not in a depression in phase space as shown e.g. in Fig. \ref{fig01}. However, the analytic simplicity motivates the approach. 
Furthermore subcases can be identified: drifting, non-drifting; oscillating, non-oscillating, as follows. Our main purpose is to provide an investigation showing a regular behavior for small $\kappa$ values, as long as $\alpha > 0$. 

\subsection{Non-drifting, non-oscillating}
\begin{figure}
\centering{
\includegraphics[width=10.5 cm]{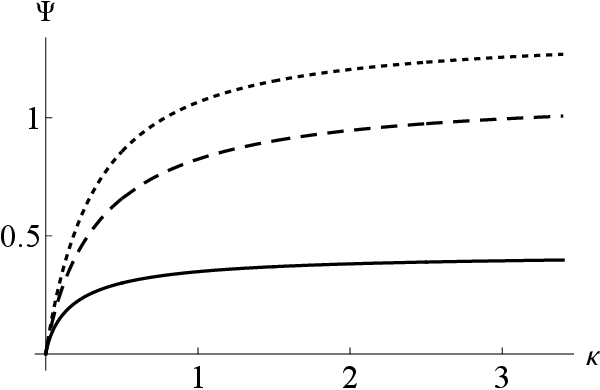}}
\caption{Solitary wave amplitude in the homogeneous trapped distribution, non-drifting and non-oscillating case as a function of $\kappa$ and different $\alpha$'s, from Eq. (\ref{ampp}). Upper, dotted line: $\alpha = 0.0$; mid, dashed: $\alpha = 0.5$; Lower, solid: $\alpha = 1.5$.}
\label{fig11}
\end{figure}

If the trapped distribution is homogeneous and non-drifting with respect to the fixed ionic background ($v_0 = 0$), one has from Eq. (\ref{ndr}) 
\begin{equation}
k_0^2 + \frac{2\,U\left(\frac{1}{2},\frac{1}{2}-\kappa,\alpha^2\kappa\right)}{\kappa\,U\left(\frac{3}{2},\frac{3}{2}-\kappa,\alpha^2\kappa\right)} - \frac{32\sqrt{2\,\Psi}}{15\,\kappa\,\sqrt{\pi\kappa}\,U\left(\frac{3}{2},\frac{3}{2}-\kappa,\alpha^2\kappa\right)} = 0 \,.
\label{fuuu}
\end{equation}
Furthermore in the non-oscillating case $k_0 = 0$ one can solve Eq. (\ref{fuuu}) as
\begin{equation}
\label{ampp}
\Psi = \frac{\pi}{2}\,\kappa\,\left[\frac{15}{16}U\left(\frac{1}{2},\frac{1}{2}-\kappa,\alpha^2\kappa\right)\right]^2 \,,
\end{equation}
which is the amplitude of the solitary wave in terms of the remaining parameters $\kappa, \alpha$ only. Figure \ref{fig11} shows the resulting amplitude. The regular behavior as $\kappa \rightarrow 0$ is apparent. A larger $\alpha$ implies a smaller solitary wave amplitude. In the non-regularized limit $\alpha \rightarrow 0$ it is possible to show that from Eq. (\ref{ampp}) one has $\Psi \rightarrow 1.38$ as $\kappa \rightarrow \infty$, which is beyond the weakly nonlinear assumption. From Fig. \ref{fig11} one also has that the $\alpha = 0$ case only admits small amplitude holes for $\kappa \ll 1$, which is in contradiction with the constraint $\kappa > 3/2$ for the non-regularized equilibrium. It is interesting to note that the weakly nonlinear condition $\Psi \ll 1$ is much better fulfilled for sufficiently high $\alpha$. Hence, such hole structures (with $\beta = 0$, non-drifting and non-oscillating) are more reliable in a RKD background. Note, however, that high $\alpha$ values limit the extent of the power laws. 

\subsection{Non-drifting, oscillating}
\begin{figure}
\centering{
\includegraphics[width=10.5 cm]{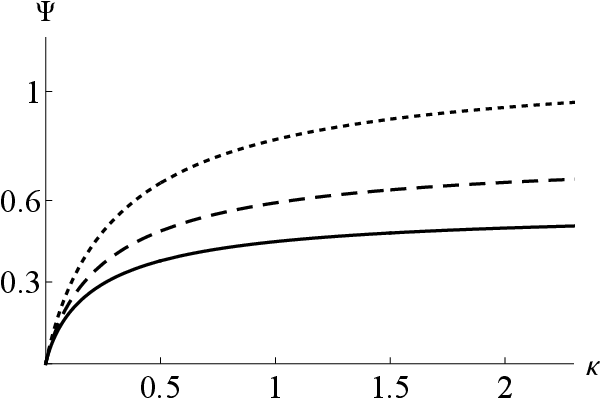}}
\caption{Wave amplitude in the homogeneous trapped distribution, non-drifting and oscillating case as a function of $\kappa$ and different wavenumbers, for $\alpha = 1.5$, from Eq. (\ref{fuuu}). Lower, solid: $k_0 = 1.0$; mid, dashed: $k_0 = 1.5$; upper, dotted line: $k_0 = 2.0$.}
\label{fig12}
\end{figure}

Allowing with $k_0 \neq 0$ for oscillating solutions one also has a regular behavior of the amplitude as $\kappa \ll 1$. In this limit, assuming $\alpha > 0$, it can be shown that Eq. (\ref{fuuu}) reduces to
\begin{equation}
\label{kkk}
k_0^2 + \frac{\sqrt{\pi}\,\alpha}{\sqrt{\kappa}} - \frac{32\,\alpha\sqrt{\Psi}}{15\,\sqrt{2\,\pi}\,\kappa} = 0 \,, \quad \kappa \ll 1 \,, \quad \alpha > 0
\end{equation}
yielding a vanishingly small amplitude as $\kappa \rightarrow 0$. Figure \ref{fig12} shows $\Psi$ from Eq. (\ref{fuuu}) as a function of $\kappa$, for $\alpha = 1.5$ and different $k_0$ values. It is found that a larger $k_0$ yields a larger  amplitude.



\subsection{Dispersion relation with $v_0 \neq 0$}
\begin{figure}
\centering{
\includegraphics[width=10.5 cm]{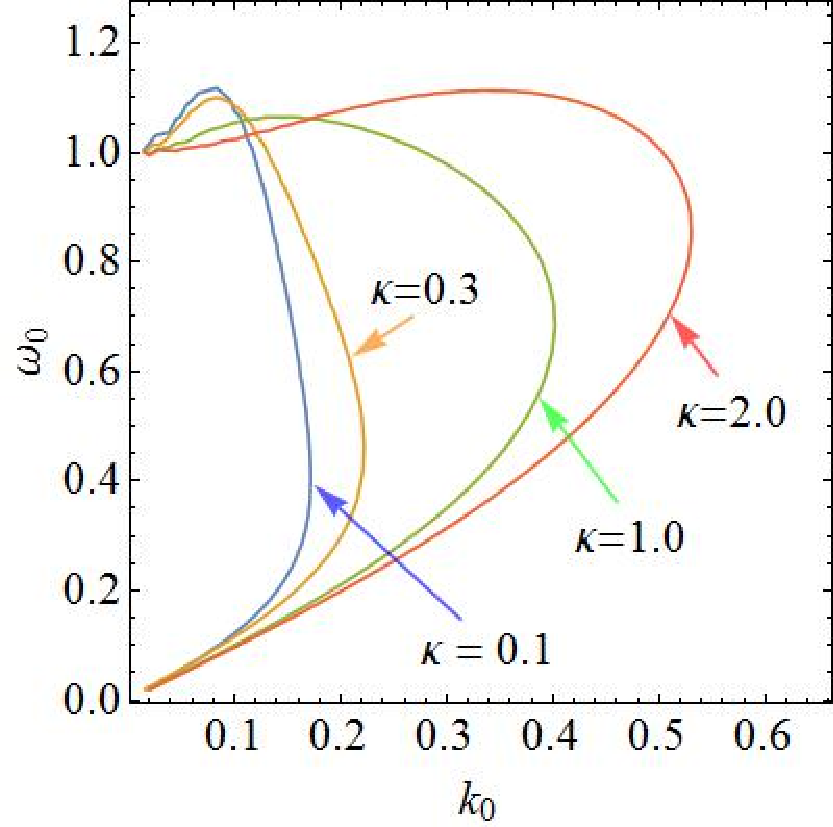}}
\caption{Dispersion relation (\ref{ldr}) with $v_0 = \omega_0/k_0$ for $\alpha = 0.1$ and $\kappa = 0.1, 0.3, 1.0, 2.0$, as indicated. }
\label{fig13}
\end{figure}

Allowing for drifting structures so that $v_0 \neq 0$, for simplicity disregarding the nonlinear term $\sim b \sqrt{\Psi}$ and still with homogeneous trapped electrons distribution ($\beta = 0$), one has from Eq. (\ref{ndr}),
\begin{eqnarray}
k_0^2 &+& \frac{2}{\kappa\,U\left(\frac{3}{2},\frac{3}{2}-\kappa,\alpha^2\kappa\right)}\,\Bigl[U\left(\frac{1}{2},\frac{1}{2}-\kappa,\alpha^2\kappa\right) + \nonumber \\
&+& \frac{v_0}{\sqrt{\pi\,\kappa}}\,P\int_{-\infty}^{\infty}\frac{ds}{s - v_0}\,e^{-\alpha^2\,s^2}\,\left(1 + \frac{s^2}{\kappa}\right)^{-\kappa-1}\Bigr] = 0 \,.
\label{ldr}
\end{eqnarray}
Setting $v_0 = \omega_0/k_0$, Eq. (\ref{ldr}) produces similar thumb curves as for holes in a Maxwellian background \citep{Schamel1986}, now adapted for the RKD. Figure \ref{fig13} show results for different small $\kappa$ values, in all cases with $\alpha = 0.1$. As usual, one has a high frequency (Langmuir) mode together with a slow electron-acoustic mode \citep{Fried} now adapted to the RKD background, where both modes coalesce in a certain point according to the parameters. As seen, the behavior is regular even for small $\kappa$ values.  At the extremal $k$ value where both modes coalesce, apparently the group velocity is infinite. As discussed in \citep{q1, q2}, at this point taking into account the nonlinear trapping the phase velocity of the hole should replace the diverging linear group velocity. 

\section{Conclusions}  
In the present paper electron holes have been discussed, for the first time in a suprathermal plasma described with a regularized kappa distribution. Unlike \citep{Haas2021}, for simplicity, here the background distribution function has no singular features. It was verified that the regularization of the standard kappa distribution avoids all divergent features the solutions for $\kappa \leq 3/2$, i.e.\ the analysis could be extended to all positive kappa values. This allows one to study plasma backgrounds that are described with velocity power-laws 
harder than $v^{-5}$ and those that exhibit an exponential cut-off, which are both observed in the solar wind.
Note also, that even for kappa values below two, which can technically be handled with an SKD, unphysical features related to a non-negligible contribution of particles at high velocities are unavoidable \citep{Scherer2019}. Their removal also requires the use of an RKD.

In terms of the hole distribution function for trapped and untrapped electrons, the number density has been evaluated yielding the pseudopotential in the weakly nonlinear limit. As a consequence, the most prominent solutions of the resulting Poisson equation have been found. Drifting, non-drifting, oscillating and non-oscillating solutions have been discussed. The linear dispersion relation has been also analyzed, yielding a $\kappa$-dependent plasma mode diagram revealing the existence of a high frequency Langmuir mode and a low frequency electron acoustic mode (Fig. \ref{fig13}). Unlike for the case of a pure power-law, i.e.\ for an SKD background, all findings based on power-laws with an exponential cut-off, i.e.\ based on an RKD background, remain regular even for very small $\kappa$ values.  The results are, therefore, relevant especially for those plasmas in a suprathermal equilibrium state with spectral index $\kappa < 3/2$, for which the SKD is not appropriate, but which are 
observed in space plasmas \citep{Gloeckler-etal-2012}.

\begin{acknowledgments}
FH acknowledges the support by Con\-se\-lho Na\-cio\-nal de De\-sen\-vol\-vi\-men\-to Cien\-t\'{\i}\-fi\-co e Tec\-no\-l\'o\-gi\-co (CNPq) and the Alexander von Humboldt Foundation for a renewed research stay fellowship.  
\end{acknowledgments}


\begin{thebibliography}{999}
\bibitem[Steinvall (2019a)]{Steinvall2019a} K. Steinvall, Y. V. Khotyaintsev, D. B. Graham, A. Vaivads, O. Le Contel, and C. T. Russell, Phys. Rev. Lett. {\bf 123}, 255101 (2019a).
\bibitem[Steinvall (2019b)]{Steinvall2019b} K. Steinvall, Yu. V. Khotyaintsev, D. B. Graham, A. Vaivads, P.-A. Lindqvist, C. T. Russell and J. L. Burch, Geophys. Res. Lett. {\bf 46}, 55 (2019b).
\bibitem[Luque (2005)]{Luque2005} A. Luque and H. Schamel, 
Phys. Rep. {\bf 415}, 261 (2005).
\bibitem[Eliasson (2006)]{Eliasson2006} B. Eliasson and P. K. Shukla, 
Phys. Rep. {\bf 422}, 225 (2006). 
\bibitem[Schamel (2015)]{Schamel2015} H. Schamel, 
Phys. Plasmas {\bf 22}, 042301 (2015).
\bibitem[Schamel (2023)]{Schamel2023} H. Schamel, Rev. Mod. Plasma Phys. {\bf 7}, 11 (2023).
\bibitem[Haas (2021)]{Haas2021} F. Haas, Phys. Plasmas {\bf 28}, 072110 (2021).
\bibitem[Aravindakshan (2018)]{w1} H. Aravindakshan, A. Kakad and B. Kakad, Phys. Plasmas {\bf 25}, 052901 (2018).
\bibitem[Aravindakshan (2020)]{w2} H. Aravindakshan, P. H. Yoon, A. Kakad and B. Kakad, Monthly Notices Royal Astron. Soc.
{\bf 497}, L69 (2020).
\bibitem[Jenab (2021)]{w3} S. M. H. Jenab, G. Brodin, J. Juno and I. Kourakis, Sci. Rep. {\bf 11}, 16358 (2021).
\bibitem[Olbert (1968)]{Olbert1968} S. Olbert, Astrophys. Space Sci. Lib. {\bf 10}, 641 (1968).
\bibitem[Lazar (2017)]{Lazar2017} M. Lazar, V. Pierrard, M. Shaaban, H. Fichtner and S. Poedts, Astron. Astrophys. 
{\bf 602}, A44 (2017).
\bibitem[Ma \& Summers (1998)]{Ma-Summers-1998} C.-Y. Ma and D. Summers, Geophys. Res. Lett. {\bf 25}, 4099 (1998).
\bibitem[Yoon (2014)]{Yoon2014} P. H. Yoon, J. Geophys. Res. {\bf 119}, 7074 (2014). 
\bibitem[Yoon et al. (2018)]{Yoon-2018} P. H. Yoon, M. Lazar, K. Scherer, H. Fichtner and R. Schlickeiser, 
Astrophys. J. {\bf 868}, 131 (2018).
\bibitem[Lazar (2021)]{Lazar2021} M. Lazar and H. Fichtner (Eds.), Astrophys. Space Sci. Lib. {\bf 464} (2021).
\bibitem[Matsumoto (1972)]{Matsumoto1972} H. Matsumoto, Ph.D. Thesis, Kyoto University, Japan (1972).
\bibitem[Lazar (2016)]{Lazar2016} M. Lazar, H. Fichtner and P. H. Yoon, Astron. Astrophys. {\bf 589}, A39 (2016).
\bibitem[Vasyliunas (1968)]{k1} V. M. Vasyliunas, J. Geophys. Res. {\bf 73}, 2839 (1968).
\bibitem[Pierrard (2010)]{k2} V. Pierrard and M. Lazar, Solar Phys. {\bf 267}, 153 (2010).
\bibitem[Hau (2007)]{k4} L.-N. Hau and W.-Z. Fu, Phys. Plasmas {\bf 14}, 110702 (2007).
\bibitem[Gloeckler et al.\ (2012)]{Gloeckler-etal-2012} G. Gloeckler. L. A. Fisk, G. M. Mason, E. C. Roelof and E. C. Stone, AIP Conf. Proc. {\bf 1436}, 136 (2012).
\bibitem[Pierrard (2022)]{Pierrard-etal-2022} V. Pierrard, M. Lazar and S. Stverak, Front. Astron. Space Sci. {\bf 9}, 892236 (2022).
\bibitem[Vasko (2017)]{Vasko2017} I. Y. Vasko, O. V. Agapitov, F. S. Mozer, J. W. Bonnell, A. V. Artemyev, V. V. Krasnoselskikh, G. Reeves and G. Hospodarsky, Geophys. Res. Lett. {\bf 44}, 4575 (2017). 
\bibitem[Scherer (2019)]{Scherer2019} K. Scherer, M. Lazar, E. Husidic and H. Fichtner, 
Astrophys. J. {\bf 880}, 118 (2019).
\bibitem[Oka et al.\ (2013)]{Oka-etal-2013} M. Oka, S. Ishikawa, P. Saint-Hilaire, S. Krucker and R. P. Lin, Astrophys. J. {\bf 764}, 6 (2013). 
\bibitem[Heerikhuisen et al.\ (2008)]{Heerikhuisen-etal-2008} J. Heerikhuisen, N. V. Pogorelov, V. Florinski, G. P. Zank and J. A. le Roux, Astrophys. J. {\bf 682} 679 (2008).
\bibitem[Zirnstein et al.\ (2017)]{Zirnstein-etal-2017}
E. J. Zirnstein, J. Heerikhuisen, G. P. Zank, N. V. Pogorelov, H. O. Funsten, D. J. McComas, D. B. Reisenfeld and N. A. Schwadron, Astrophys. J. {\bf 836} 238 (2017).
\bibitem[Fisk \& Gloeckler (2012)]{Fisk-Gloeckler-2012} L. A. Fisk and G. Gloeckler, Space Sci. Rev. {\bf 173} 433 (2012).
\bibitem[Scherer (2017)]{Scherer2017} K. Scherer, H. Fichtner and M. Lazar, Europhys. Lett. {\bf 120}, 50002 (2017). 
\bibitem[HanThanh (2022)]{HanThanh2022} L. Han-Thanh, K. Scherer and H. Fichtner, Phys. Plasmas {\bf 29}, 022901 (2022).
\bibitem[Liu (2020)]{Liu2020} Y. Liu, AIP Advances {\bf 10}, 085022 (2020).
\bibitem[Abramowitz (1972)]{Abramowitz1972} M. Abramowitz and I. A. Stegun, {\it Handbook of Mathematical Functions, with Formulas, Graphs, and Mathematical Tables} (10 ed.), 	United States Department of Commerce, National Bureau of Standards (1972).
\bibitem[Summers (1991)]{Summers} D. Summers and R. M. Thorne, 
Phys. Fluids B {\bf 3}, 1835 (1991).
\bibitem[Podesta (2005)]{Podesta} J. J. Podesta, 
Phys. Plasmas {\bf 12}, 052101 (2005).
\bibitem[Schamel (1972)]{Schamel1972} H. Schamel, 
Plasma Phys. {\bf 14}, 905 (1972).
\bibitem[Schamel (2018)]{kkk} H. Schamel, N. Das and P. Borah, Phys. Lett. A {\bf 382},  168 (2018).
\bibitem[Schamel (1986)]{Schamel1986} H. Schamel, 
Phys. Rep. {\bf 140}, 161 (1986).
\bibitem[Schamel (2012)]{Schamel2012} H. Schamel, 
Phys. Plasmas {\bf 19}, 020501 (2012). 
\bibitem[Schamel (2000)]{sch} H. Schamel, Phys. Plasmas {\bf 7}, 4831 (2000).
Phys. Plasmas {\bf 21}, 092103 (2014).
\bibitem[Fried (1961)]{Fried} B. D. Fried and R. W. Gould, Phys. Fluids {\bf 4}, 139 (1961). 
\bibitem[Schamel (2013)]{q1} H. Schamel, Phys. Plasmas 20, 034701 (2013).
\bibitem[Valentini (2012)]{q2} F. Valentini, D. Perrone, F. Califano, F. Pegoraro, P. Veltri, P. J. Morrison and T. M. O’Neil,
Phys. Plasmas 19, 092103 (2012).
\end{thebibliography}
\end{document}